\documentstyle[12pt,epsfig]{article}

\begin{document}
\newcommand {\be}{\begin{equation}}
\newcommand {\ee}{\end{equation}}
\newcommand {\ba}{\begin{eqnarray}}
\newcommand {\ea}{\end{eqnarray}}
\newcommand {\bea}{\begin{array}}
\newcommand {\cl}{\centerline}

\newcommand {\eea}{\end{array}}

\renewcommand {\thefootnote}{\fnsymbol{footnote}}
\vskip .5cm
\newcommand {\prl} {Phys. Rev. Lett.}
\newcommand {\dar} {CITATION = HEP-PH}
\newcommand {\plb} {Phys. Lett. B}
\newcommand {\prd} {Phys. Rev. D}
\renewcommand {\thefootnote}{\fnsymbol{footnote}}

\def \a'{\alpha'}
\baselineskip 0.65 cm
\begin{flushright}
IPM/P-2004/073 \\
hep-ph/0411358 \\
\today
\end{flushright}
\begin{center}
{\Large{\bf
Effects of the neutrino $B$-term on the Higgs mass parameters and
electroweak symmetry breaking}
}
{\vskip 0.5 cm}

{\bf Yasaman  Farzan\\}
{\vskip 0.5 cm}
Institute for Studies in Theoretical Physics and Mathematics (IPM)\\
P.O. Box 19395-5531, Tehran, Iran\\
The Abdus Salam International Centre for
Theoretical Physics (ICTP)
\\
Strada Costiera 11, 34100 Trieste, Italy\\

\begin{abstract}
To embed the seesaw mechanism in the MSSM, two or three
right-handed
neutrino
supermultiplets, $N_i$, have to be added to the model. In the presence
of these new superfields,
the soft supersymmetry breaking potential  includes
 a lepton number violating mass term
which is known as the neutrino $B$-term, $M
B_\nu\tilde{N}\tilde{N}/2$. In this paper, we study the effects of
$B_\nu$ on the Higgs mass parameters. Using the condition for the
electroweak symmetry breaking, we derive an upper bound on
 $B_\nu $ which,  for some range of parameters, is
  two orders of magnitude stronger than the
previous bounds. We also propose a simple model in which it is
natural to have large values of $B_\nu$ while the rest of the
supersymmetry breaking terms are at the TeV scale or smaller.
\end{abstract}
\end{center}

\section{Introduction}
The Standard Model (SM) of the particle physics so far has had an enormous
success. However, there are two cases that the SM fails to provide
an explanation: the neutrino oscillation data and the stability of the
electroweak scale.

There is a consensus among the physicists
 that to accommodate the solar, atmospheric and reactor data,
neutrinos have to be massive. On the other hand, from the
cosmological data as well as the studies of the end-point of the Tritium
beta
decay, we know that  the masses of neutrinos are very tiny:
$m_\nu/m_e<10^{-6}$. One way to attribute a tiny but nonzero mass to
neutrinos
is the famous seesaw mechanism which adds heavy right-handed neutrinos to
the SM.

To stabilize the electroweak symmetry breaking scale against radiative
corrections, several models have been developed among which
supersymmetry is arguably the most popular one.
The simplest version of supersymmetry is the Minimal Supersymmetric
Standard
Model (MSSM).

To build a model that incorporates the both mechanisms, we can augment
the
MSSM to include three heavy right-handed neutrino
supermultiplets, $N_i$.
In the presence of the right-handed
neutrinos the superpotential will be
\be
W=Y_\ell^{i j} \epsilon_{\alpha \beta}H_d^\alpha
l^c_{Ri}  L_j^\beta+ Y_\nu^{i j}
\epsilon_{\alpha \beta}H_u^\alpha N_{i}
L_j^\beta+\frac{1}{2}M_{ij} N_{i} N_{j}+(\mu H_u H_d+{\rm H.c.}),
\ee
where $L_j^\beta$ is the supermultiplet
corresponding to the
 doublet $(\nu_{Lj},~ l_{Lj})$. Since the quark sector is irrelevant for
our discussion, we have omitted  the terms involving  quarks in the
above
formula.
Without loss of
generality, we can rephase and rotate the
fields to make the matrices
$Y_\ell^{ij}$ and $M_{ij}$  real and diagonal:
$Y_\ell^{ij}={\rm diag}(Y_e, Y_\mu, Y_\tau)$ and
$M^{ij}={\rm
diag}(M_1, M_2, M_3)$. In this basis, $Y_\nu$ can
have off-diagonal and complex elements.
We will assume that $M_1$, $M_2$ and $M_3$ are much larger than
$m_{susy}$.

%It is well-known that non-zero lepton-flavor-violating slepton
%mass terms in the soft supersymmetry breaking Lagrangian ($m_{\alpha
%\beta}^2
%\tilde{L}_\alpha^\dagger \tilde{L}_\beta  \ \
%\alpha \ne \beta$) can give
%rise  to rare decays such as ($\mu \to \gamma e$),
%($\tau \to \gamma e$) and ($\tau \to \gamma
%\mu$). One way to avoid  flavor changing neutral current (FCNC) effects
%is to choose the
%off-diagonal mass terms to be small.
For simplicity, in this paper we  work in the framework of
 the mSUGRA for which at the GUT
scale, the soft supersymmetry breaking terms are flavor
blind; that is  at the GUT scale
 \ba
-{\cal
L}_{soft}
&=&
m_0^2(\tilde{L}_{L\alpha}^\dagger\tilde{L}_{L\alpha}+
\tilde{l}_{R\alpha}^\dagger \tilde{l}_{R\alpha}+
\tilde{N}_{\alpha}^\dagger
\tilde{N}_{\alpha}+H_d^\dagger
H_d+ H_u^\dagger H_u \label{soft})
+
 \frac{1}{2} m_{1/2}(\tilde{B}^\dagger \tilde{B}+
\tilde{W^a}^\dagger \tilde{W^a}) \cr &+& (b^0_H  H_d H_u +h.c.)+
A_\ell^{ij}\epsilon_{\alpha \beta}
H_d^\alpha\tilde{l}_{Ri}^\dagger \tilde{L}_{Lj}^\beta+
A_\nu^{ij}\epsilon_{\alpha\beta} H_u^\alpha \tilde{N}_{i}
\tilde{L}_{Lj}^\beta \cr &+& (\frac{1}{2} B_\nu M_i
\tilde{N}^i\tilde{N}^i+h. c.), \ea where $A_\ell=a_0 Y_\ell$ and
$A_\nu=a_0 Y_\nu$. The generalization to a more general
supersymmetry breaking scheme is straightforward. Due to the
radiative corrections, at the electroweak scale the masses of
different scalars; in particular, the masses of $H_u$ and $H_d$
($m^2_{H_u}$ and $m^2_{H_d}$) will be different.

%The last term in (\ref{soft}) is the neutrino $B$-term which breaks
%lepton
%number by two units. For small values of $B_\nu$ ($\sqrt{B_\nu M}\sim
%m_{susy}$), the phase of neutrino $B$-term [more precisely,
% ${\rm arg}(A_\nu M B_\nu^*Y_\nu^*)$] can give rise to CP-violation that
%may account for the Baryon asymmetry of the Universe \cite{kashti}.
If  $|B_\nu|$
is large  ($B_\nu\gg m_{susy}$), several
interesting phenomena can occur. For example, as discussed in \cite{g-h},
a large $B_\nu$ can give rise to $\tilde{\nu}-\bar{\tilde{\nu}}$
oscillation in
the linear
colliders. Moreover, large $B_\nu$ can give a radiative correction to
the masses of left-handed neutrinos \cite{g-h}.
The neutrino $B$-term also gives a correction to the mass matrix of
left-handed sleptons which, in principle, can be Lepton Flavor Violating
(LFV),  inducing LFV rare decays \cite{khodam}.
Finally, the neutrino $B$-term gives a correction to the $A$-term of the
charged leptons, $A_\ell$, proportional to $B_\nu$. As it is
well-known, an imaginary $A_\ell$ can
induce  Electric Dipole Moment (EDM) for the charged leptons so
this way imaginary $ B_\nu$ induces EDMs for the charged leptons.

 The strong upper bounds on the neutrino mass and branching
ratios of LFV rare decays and the EDM of
the electron give bounds on different combinations of $B_\nu$, $Y_\nu$ and
$M$. However, as we shall discuss, these bounds are not enough to
restrict the value of $B_\nu$, independently of the values of other
parameters.

In this paper, we will show that  $B_\nu$ can also give radiative
corrections to the Higgs mass parameters. The corrections to $m_{H_u}^2$
and $b_H$ turn out to be finite and of order of $B_\nu
m_{susy}Y_\nu^2/(16\pi^2)$.
Assuming that there is no significant cancelation between the tree
level values of $b_H$ and $m_{H_u}^2$ and the loop corrections, the
condition for the electroweak symmetry breaking puts  a strong bound
on
the neutrino $B$-term: $B_\nu Y_\nu^2/(4 \pi)^2<m_{susy}/\tan \beta$.

In section 2, we suggest a simple model which makes a large neutrino
$B$-term naturally possible.
In section 3, we study the effects of the neutrino $B$-term on the
Higgs mass parameters and derive a bound on $B_\nu$.
In section 4, we review the effects of $B_\nu$ on different observables
and discuss how the bound we find in section 3 complements our knowledge
of this sector.
Finally, in section 5, we summarize our conclusions.

\section{Theoretical framework for a large $B_\nu$}
In this section, we first review what  the ``natural" size
of the neutrino $B$-term in the context of the mSUGRA is. Then, we suggest
a model in which larger values
of $B_\nu$ are acceptable.

In the context of the mSUGRA,
the soft supersymmetry breaking terms originate from the interaction of
a chiral superfield $S$ with the super-potential:
\be
\int d^2 \theta S(\theta) W(\theta).
\ee
The scalar and $F$-components of $S$ develop vacuum expectation values
$\langle S \rangle=1 +F_S \theta^2$ and $\langle F_S \rangle$
determines the scale of  the soft supersymmetry breaking terms.
Within this model we expect $B_\nu \sim a_0 \sim m_{susy}$.
Remember that we have parameterized the neutrino $B$-term as $M B_\nu
\tilde{N}\tilde{N}/2$ so, in this model, we expect
$\sqrt{B_\nu M}\gg m_{susy}$.

%terms such as
%$$\int(1+A_\Phi \theta^2+\bar{A}_\Phi\bar{\theta}^2+B_\Phi\theta^2
%\bar{\theta}^2)
%\Phi^\dagger \Phi d^4\theta$$
%are the origins of the soft
%supersymmetry breaking potential.
%Taking $\Phi$ to be the right-handed neutrino supermultiplet, we
%conclude $B_\nu\sim A_\nu \sim A_N$. In principle, we can take $A_N\gg
%m_{susy}(\sim m_0,a_0)$. However, this is not a plausible model because
%there is no symmetry protecting other supersymmetry breaking terms
%($m^2_0$,
%$m_{\tilde{\chi}}$, $b_H$ and ect.) from being of order of $A_N$.
%The natural size for $A_N$ in this model is $m_{susy}$; that is
%$B_\nu\sim
%A_\nu\sim
%m_{susy}$.
% and as a result, the soft leptogenesis
%mechanism cannot work \cite{kashti}.
%To have a successful soft leptogenesis, there has to be
%a mechanism to cancel the leading contribution to $B_\nu$ \cite{thappa}.

In this paper, we are more interested in large values of
 $B_\nu$ ($B_\nu Y_\nu^2/16 \pi^2 \sim m_{susy}$) and we
propose a model that can provide us with large values of $B_\nu$.

Let us assume that besides $S$ which couples to the lepton number
conserving
part of the superpotential,
there is  a spurion field, $X$, that carries
lepton number equal to two. Then we can write the following term in the
superpotential
\be \label{spurion}\int d^2 \theta \lambda_i X N_i N_i. \ee
However, terms such as $\int d^2 \theta X H_u H_d$ are forbidden
by lepton number conservation. Moreover terms such as $\int X^\dagger X
\Phi^\dagger \Phi d^4 \theta$ in the K\" ahler potential are suppressed
by powers of $M_{pl}^{-1}$.
Let us assume that the self-interaction of the hidden sector is such
that both the scalar- and $F$-components of $X$ develop nonzero vacuum
expectation
values.
The vacuum expectation values  of the components of $X$ break the lepton
number symmetry of the model. The vacuum expectation value of the scalar
component of $X$, $\langle \tilde{X}\rangle$, corresponds to the Majorana
mass
term of the right-handed neutrinos while the  vacuum expectation value of
the $F$-component, $\langle F_X \rangle$, gives  the neutrino $B$-term.
With our parameterization of the neutrino $B$-term,
\be
B_\nu ={\langle F_X\rangle \over \langle \tilde{X} \rangle}.
\ee
Both
$\langle F_X\rangle$ and  $\langle \tilde{X} \rangle$ can be large,
giving rise to large right-handed neutrino masses and $B_\nu$ while other
supersymmetry breaking terms, being given by $\langle F_S \rangle$, are
at TeV scale or smaller. Notice that in this model, in the basis
that the mass matrix of the right-handed neutrinos is
real diagonal, the neutrino $B$-term is also diagonal so the
parameterization that we are using for the neutrino $B$-term is
appropriate.

Notice that this model is very similar to the singlet Majoron
model \cite{majoron},
with the difference  that here $\langle F_X \rangle$ is also nonzero.

If the $X$ field is very heavy, it has to decouple from the theory
\cite{carrozone}. In particular although,  the term in  (\ref{spurion})
 gives rise to a quartic right-handed sneutrino
self-interaction term ($|\sum_i \lambda_i \tilde{N_i}
\tilde{N_i}|^2$), based on the decoupling theorem we expect that
there has to be a term (for example, coming from the mass term of
$S$) canceling this effect. Even in the absence of such
a cancelation, as far as $|B_\nu|<|M|$, right-handed sneutrinos do
not develop  non-zero vacuum expectation values and the scheme of
the electroweak symmetry breaking will be similar to the MSSM. In
the next section, we will take an agnostic approach and will not
care about the origin of $B_\nu$.
\section{Effects of the neutrino $B$-term on the Higgs mass
parameters}
In this section, we study the effects of a large $B_\nu$ on the Higgs mass
parameters and derive bounds on its value from the fulfillment of the
electroweak
symmetry breaking condition.
Then we compare this bound with the limit derived by considering the
effect of $B_\nu$ on the neutrino mass.

As we shall see, the effects of the neutrino $B$-term on
the Higgs mass parameters are finite so they cannot be derived by
integrating the
renormalization group equations. However, since the momentum propagating
in the loop is of order of $M_i$, to calculate the effect
precisely, we have to
insert the values of the relevant parameters at the energy  scale of
the right-handed neutrinos,
$M_i$,
instead of those at the GUT scale.
The running of
the $A$-terms, Yukawa coupling and the right-handed (s)neutrino masses
have been discussed in detail in the literature \cite{rge}.
The running of the neutrino $B$-term is given by
\be \label{bruns}
16 \pi^2 \mu {d~B_\nu M \over d~\mu}=2A_\nu Y_\nu^\dagger M +2M
Y_\nu^* A_\nu^T
-B_\nu M Y_\nu^\dagger Y_\nu
- Y_\nu^T Y_\nu^* B_\nu M.
\ee
Assuming $|A_\nu| \ll |B_\nu|$, the first term is negligible.
The running of the mass matrix of the right-handed neutrinos
is
\be
\label{mruns}
16 \pi^2\mu {d~M \over d~\mu}=  -M Y_\nu^\dagger Y_\nu- Y_\nu^T Y_\nu^* M
\ee
Comparing Eqs. (\ref{bruns}) and (\ref{mruns}),  we observe that,
 up to a correction of
${\cal O}(A_\nu/B_\nu)$,  the neutrino $B$-term remains proportional to
the mass matrix of the right-handed neutrinos.
\begin{figure}
\begin{center}
%\vskip -3.2cm
\hskip -7.0cm
\parbox[c]{3.5in}
{\mbox{
\qquad
\epsfig
{file=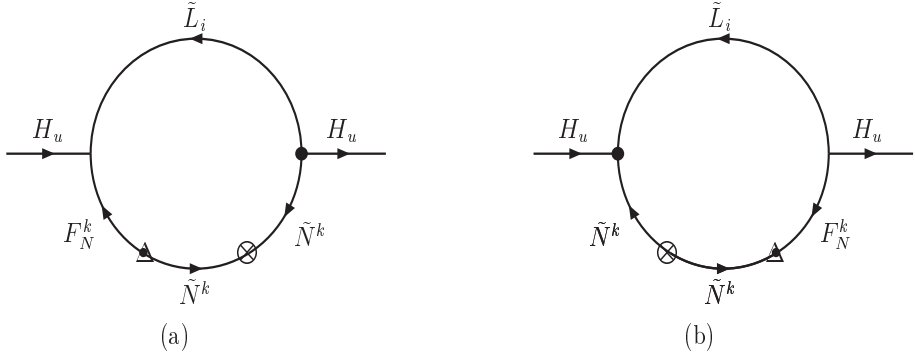,
bb=75 578 528 744, clip=true,
width=5 in, height=2 in}}}
\end{center}
\vskip -0.8 cm
\caption{Diagrams contributing to $m_{H_u}^2$. $F_N^k$
represents the auxiliary field associated with the right-handed neutrino,
$N_k$. The $A_\nu$ vertices are marked with black circles. The neutrino
$B$-term and $M$ insertions are indicated by $\otimes$ and $\Delta$,
respectively.}
 \label{bh2}
\end{figure}

Diagrams shown in Fig. \ref{bh2} give a correction
to $m_{H_u}^2$ which is equal to
\be
-i \Delta m_{H_u}^2=2\sum_k \int { M_k^2 {\rm Re}[B_\nu \sum_i
(Y_\nu)_{ki} (A_\nu^*)_{ki}]
\over k^2(k^2-M_k^2)^2} {d^4 k \over (2\pi)^4}
=-i 2\sum_{k,i} {\rm Re}{\left[ B_\nu {\rm Tr}(Y_\nu
A_\nu^\dagger)\right]\over 16 \pi^2}.
\label{mhucorrection}
\ee

\begin{figure}
\begin{center}
%\vskip -3.2cm
\hskip -7.0cm
\parbox[c]{3.5in}
{\mbox{
\qquad
\epsfig
{file=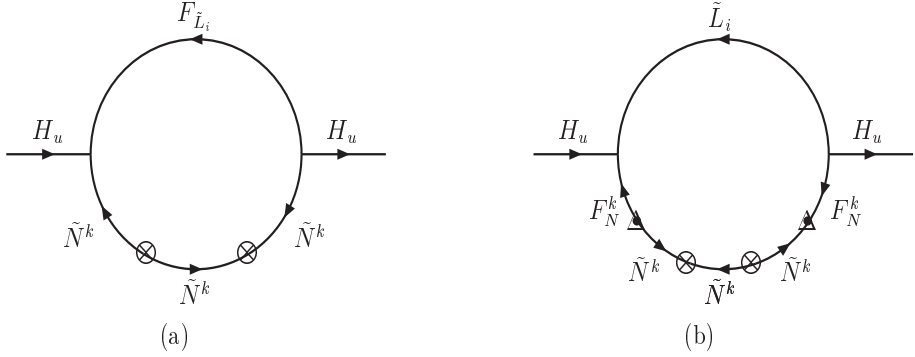,
bb=75 578 528 744, clip=true,
width=5 in, height=2 in}}}
\end{center}
\vskip -0.8 cm
\caption{Diagrams proportional to $|B_\nu|^2$ contributing to $m_{H_u}^2$.
$F_N^k$ and $F_{\tilde{L_i}}$
represent the auxiliary fields associated with
$N_k$ and $\tilde{L_i}$, respectively.
The neutrino
$B$-term and $M$ insertions are indicated by $\otimes$ and $\Delta$,
respectively. These diagrams cancel each other.
}
 \label{nob2}
\end{figure}
 Notice that this result is independent of $M_k$.
 That is because in the diagram (\ref{bh2})
 there are two vertices proportional to $M_k$  (the $B$-term,
 $MB_\nu \tilde{N}\tilde{N}$, and  $\langle F_N N\rangle \propto M$)
  which cancel the factors of
 $M$ in the denominator. Had we defined the
 $B$-term as $B_\nu \tilde{N}\tilde{N}$ (with $B_\nu$
 being of dimension of two), the result
 would have  been inversely proportional to $M$.

\begin{figure}
\begin{center}
%\vskip -1 cm
\hskip -5.0cm
\parbox[c]{3.5in}
{\mbox{
\qquad
\epsfig
{file=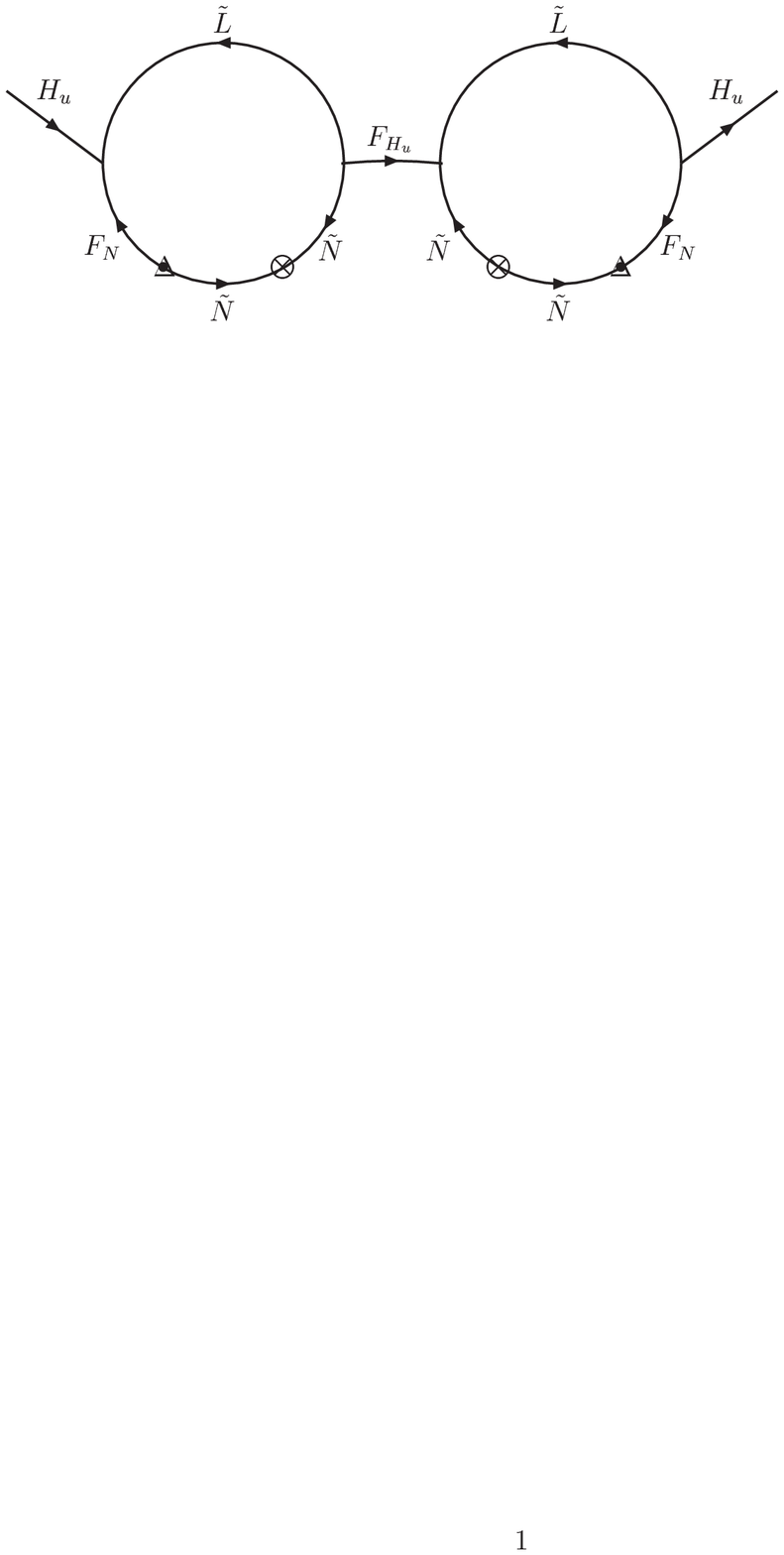,
bb=75 578 528 744, clip=true,
width=5 in, height=2 in}}}
\end{center}
\vskip -0.8 cm

\caption{
The ``eyeglasses-diagram."  The neutrino
$B$-term and $M$ insertions are indicated by $\otimes$ and $\Delta$,
respectively. This
diagram gives a correction to
$m_{H_u}^2$
equal
to $|B_\nu|^2\left[ {\rm Tr}
\{Y_\nu^\dagger Y_\nu \}/16 \pi^2 \right]^2$. Diagrams with a different
topology have a different dependence on $Y_\nu$ and cannot cancel the
effect of the ``glasses-diagram." This demonstrates that the cancelation
at one-loop level (see Fig \ref{nob2}) is completely accidental.}
 \label{glasses}
\end{figure}

 At one loop level, there is no correction proportional
to $|B_\nu|^2$ because the corresponding diagrams (shown in Fig.
\ref{nob2}) cancel each other. The cancelation seems to be
accidental because there is no symmetry forbidding such a
contribution. At the two-loop level, there is only one diagram
proportional to $\left[{\rm Trace}[Y_\nu^\dagger Y_\nu]\right]^2$.
This diagram (which we will call the ``eyeglasses" diagram) is
shown in Fig. \ref{glasses}.
 Diagrams with a
different topology are
proportional to other
combinations of the Yukawa couplings ($Y_{ki}$  and $Y_{ki}^*$) and
cannot
cancel the effect of the eyeglasses diagram. So, at the two-loop level,
there is a non-zero
effect proportional to $|B_\nu|^2$.
Performing the full two-loop calculation is beyond the scope of
this paper. The two-loop effect becomes significant  only if $B_\nu
Y_\nu^2/(16\pi^2)> m_{susy}$ but as we shall see, such a possibility
is quite unlikely.

\begin{figure}
\begin{center}
%\vskip -3.2cm
\hskip -7.0cm
\parbox[c]{3.5in}
{\mbox{
\qquad
\epsfig
{file=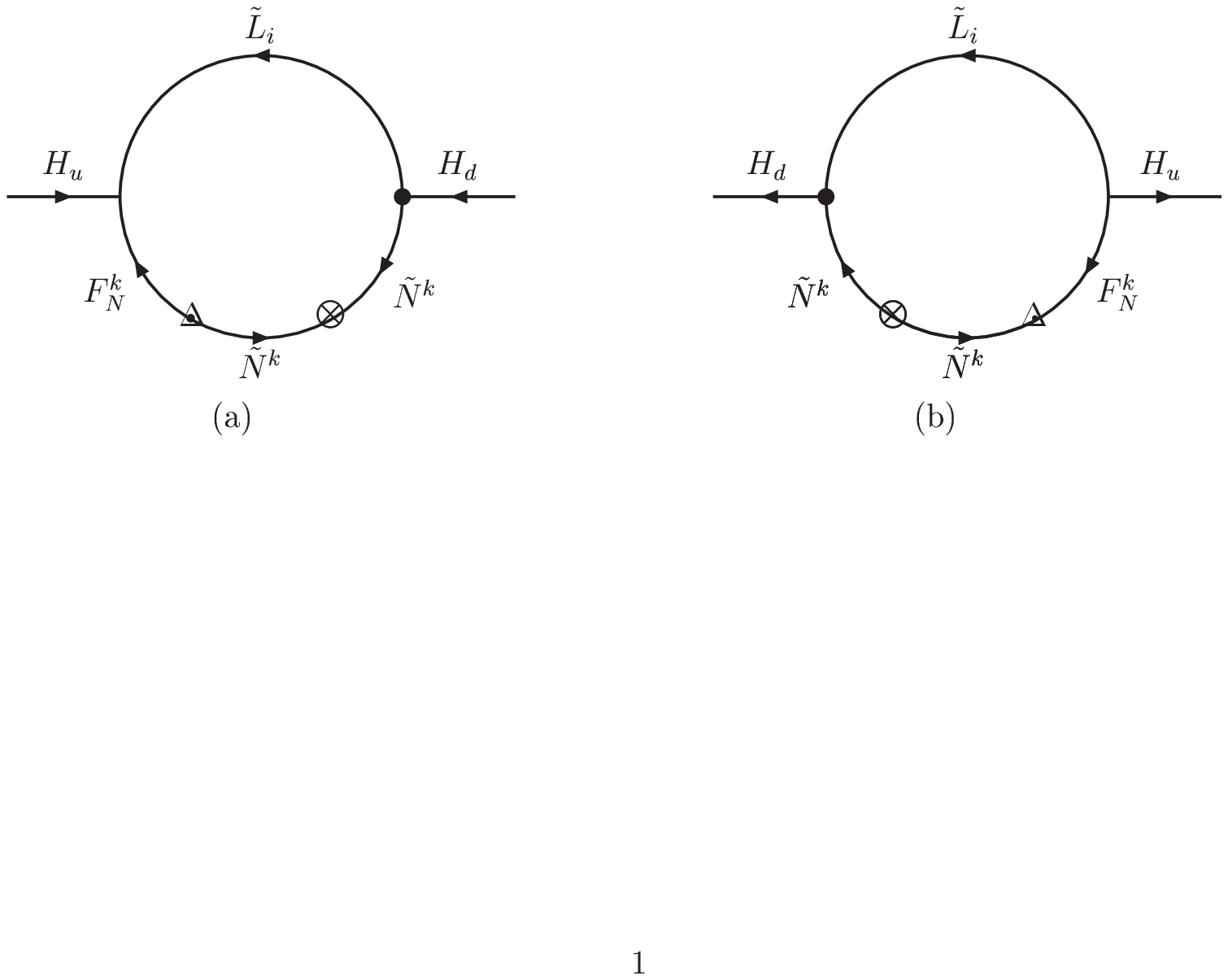,
bb=68 273 522 445, clip=true,
width=5 in, height=2 in}}}
\end{center}
\vskip -0.8 cm

\caption{
 Diagrams contributing to
 the Higgs $B$-term. $F_N^k$ represents the auxiliary
field associated with
the right-handed neutrino, $N_k$. The  two vertices marked
with black circles are given by $\mu Y_\nu^\dagger$ and $\mu^* Y_\nu$.
The neutrino
$B$-term and $M$ insertions are indicated by $\otimes$ and $\Delta$,
respectively.}
\label{bbh}
\end{figure}

The presence of a large neutrino $B$-term also
induces non-negligible corrections to $b_H$ as it is shown in
Fig \ref{bbh}.
The correction is finite and equal to
\be \label{bhcorrection}
-i \Delta b_H=-B_\nu \sum_k \int {M_k^2 {\rm Tr}\left[ (Y_\nu)_{ki}
(Y_\nu^*)_{ki} \right] \mu \over k^2(k^2-M_k^2)^2} {d^4k \over (2
\pi)^4}
={i B_\nu \mu {\rm Tr}\left[ Y_\nu Y_\nu^\dagger\right] \over
(4\pi)^2}.\ee

By dimensional analysis we can show that any correction due to $B_\nu$
to the quartic Higgs interaction is suppressed by $B_\nu /M$ which
is negligible. The contribution to the cubic Higgs term is also zero.
So the  potential of $H_u^0$ and $H_d^0$ is (see {\it e.g.,}
\cite{martin})
\be
V=(|\mu|^2+m_{H_u}^2 +\Delta
m_{H_u}^2)|H_u^0|^2+(|\mu|^2+m_{H_d}^2)|H_d^0|^2
 \label{V}
\ee
$$+[(b_H+\Delta b_H) H_u^0H_d^0+{\rm H. c.}]+{g^2+g'^2\over
8}(|H_u^0|^2-|H_d^0|^2
)^2.
$$
Requiring $m_Z^2=(g^2+g'^2)(\langle H_u\rangle^2+\langle
H_d\rangle^2)/2$ and $\partial V/\partial H_u^0=\partial
V/\partial H_d^0=0,$ we find \be |\mu|^2+m_{H_d}^2=\left|
b_H+\Delta b_H\right| \tan \beta -(m_Z^2/2)\cos 2\beta\label{con1}
\ee and \be |\mu|^2+m_{H_u}^2+\Delta m_{H_u}^2=\left| b_H+\Delta
b_H \right| \cot \beta +(m_Z^2/2)\cos 2\beta\label{con2} \ee where
$\tan \beta =\langle H_u\rangle /\langle H_d \rangle $. Assuming
$|\mu|^2\sim m_{H_u}^2 \sim m_{susy}^2$, (\ref{con1}) gives \be
\label{bound} \left| b_H-B_\nu \mu {{\rm Tr} [Y_\nu Y_\nu^\dagger
] \over 16 \pi^2} \right|\sim m_{susy}^2/\tan \beta. \ee From the
LEP data \cite{lep}, we know that $\tan \beta>2 $ and the data
favors large values of $\tan \beta$ ($\tan \beta >10$). Based on
naturalness condition, it seems quite unlikely that $b_H$ and
$\Delta b_H$ cancel each other, so we expect that \be B_\nu
Y_\nu^2/(16 \pi^2) <m_{susy}/\tan \beta. \label{dis} \ee Notice
that if $Y_\nu\ll 1$, $B_\nu$ can be still several orders of
magnitude larger than $m_{susy}$.
\section{Bounds on $B_\nu$ from different observables}
In this section, we review the effects of $B_\nu$ on different
observables and  discuss how the bound we have found in the
previous section completes our knowledge of this sector.

As it is discussed in \cite{g-h}, $B_\nu$ gives a correction to
the neutrino mass equal to \footnote{Note that, within this model,
there is another contribution to $m_\nu$; namely, the standard
seesaw effect given by $Y_\nu^T (\langle H_u \rangle^2 /M)Y_\nu$.
In general, the neutrino mass is the sum of the two contributions
and, a priori, we do not know which one is dominant.} \be -{g^2
\over 32 \pi^2 \cos^2 \theta_W}{2 B_\nu \over m_{\tilde{\nu}}}
Y_\nu^T {\langle H_u \rangle^2 \over M} Y_\nu \sum_j
f(y_j)|Z_{jZ}|^2,
 \label{mnu}
\ee where  $$ f(y_j)={\sqrt{y_j}[y_j-1-\log (y_j)] \over
(1-y_j)^2},$$  $y_j \equiv
m_{\tilde{\nu}}^2/m_{\tilde{\chi}_j^0}^2$ and $Z_{jZ}\equiv
Z_{j2}\cos \theta_W-Z_{j1}\sin \theta_W$ is the neutralino mixing
matrix element that projects out the $\tilde{W^0}$ eigenstate from
the $j$th neutralino. Using  (\ref{mnu}), authors of Ref.
\cite{g-h} have concluded that \be \label{yuval} B_\nu <10^3
m_{susy}. \ee For $Y_\nu\sim 1$ (which is a value
suggested by the unification models), the bound that we found in the
previous section [see Eq. (\ref{dis})] is two orders of magnitude
stronger. Even if $b_H\sim m_{susy}$ \footnote{Note that, for $\tan \beta > 10$, if $b_H\sim m_{susy}
$, there has to be a fine-tuned cancelation between the
two terms in  Eq. (\ref{bound}) which seems unnatural.}, the bound from  the electroweak
symmetry breaking [$B_\nu\stackrel{<}{\sim}16\pi^2 m_{susy}/({\rm Tr}\{Y_\nu
Y_\nu^\dagger \} )$] is more restrictive.
 At first
sight, it may seem that unlike the bound from the electroweak
symmetry breaking, the bound in Eq.
(\ref{yuval}) is independent of $Y_\nu$ but the fact is that  to
derive Eq. (\ref{yuval}), it has been assumed that the
$B_\nu$-induced contribution to $m_\nu$ is sub-dominant and the
main contribution to $m_\nu$ is given by the standard tree-level
seesaw formula: $m_\nu \simeq Y_\nu^T(\langle H_u\rangle
^2/M)Y_\nu $. However, in principle, $Y_\nu^T(\langle H_u\rangle
^2/M)Y_\nu$ can be much smaller than $m_\nu$ and  the dominant
contribution to the neutrino mass can be the $B_\nu$-induced
one-loop effect given in Eq. (\ref{mnu}). In this case, we can
write  \be \sum_k (Y_\nu)_{ki}(Y_\nu)_{kj}({\rm GeV} / M_k)
B_\nu/m_{susy}\sim 10^{-12} (m_{ij}/0.1 \ \ {\rm
eV})\sin^{-2}\beta. \label{b=m}\ee In general Yukawa couplings are
complex numbers and as a result different terms in the summation
can cancel each other, allowing very large values of $B_\nu$.
 Let us assume that there is not any
significant cancelation between the different terms in Eq.
(\ref{b=m}). In this case, if $M<10^{14} \ \ {\rm GeV}/\tan
\beta$, the bound from the neutrino mass on $B_\nu$ will be
stronger than (\ref{bound}); otherwise, the bound from the electroweak
symmetry breaking will be more restrictive. In general, these two bounds
  are complementary because, a priori, we  do not know the
values of $M_k$. Moreover the different terms contributing to  Eq.
(\ref{b=m}) can cancel each other; that is while,  in the case of
the bound from the electroweak symmetry breaking, the combination
of the Yukawa couplings that appears  is $\sum_{ki}
|(Y_\nu)_{ki}|^2$ and therefore such a cancelation is not
possible.

 As it is discussed in Ref. \cite{khodam}, $B_\nu$
also gives a contribution to the mass matrix of the left-handed sleptons:
\be \label{lfv} \Delta m_{\tilde{L}}^2=-2 {\rm Re}( B_\nu^* a_0)
{Y_\nu^\dagger Y_\nu
\over 16 \pi^2}.\ee
The off-diagonal elements of this matrix are LFV and, in principle,
can induce LFV rare decays. As it is well-known
there are strong bounds on the branching ratio of the rare LFV decays
\cite{pdg,belle}.
 There are two possibilities to satisfy these bounds: (i) the off-diagonal
elements of the matrix in Eq. (\ref{lfv}) are much lower than $m_{susy}$;
(ii) $B_\nu Y_\nu^2/16 \pi^2\gg m_{susy}$ and therefore $\tilde{L}$
becomes so heavy that the relevant diagrams become suppressed.
The bound we found in the previous section rules out the second
possibility. Then, the bounds on the branching ratio of the rare LFV
decays
 [${\rm Br}(\mu \to e \gamma)<1.2 \times 10^{-11}$ \cite{pdg},
${\rm Br}(\tau \to e \gamma)<2.7 \times 10^{-6}$ \cite{pdg}, ${\rm
Br}(\tau \to \mu \gamma)<3.1\times 10^{-7}$ \cite{belle}] give
strong bounds on the off-diagonal elements of the matrix \be {{\rm
Re}(a_0 B_\nu^*) \over (4 \pi)^2 m_{susy}^2}(Y_\nu^\dagger
Y_\nu)_{\mu e}< {\rm few}\times 10^{-4} \ \ \ \ \ \ \ {{\rm
Re}(a_0 B_\nu^*) \over (4 \pi)^2 m_{susy}^2}(Y_\nu^\dagger
Y_\nu)_{\tau e}<0.1 \ee and \be {{\rm Re} (a_0 B_\nu^*) \over (4
\pi)^2 m_{susy}^2}(Y_\nu^\dagger Y_\nu)_{\tau \mu}< {\rm
few}\times 10^{-2}.\ee Here, we have implicitly assumed that
different possible LFV effects do not cancel each other. Note that
these bounds are stronger than the bound we found in the previous
section. However, these bounds apply only to the off-diagonal
elements of the matrix $Y_\nu^\dagger Y_\nu$. In principle, the
off-diagonal elements of $Y_\nu^\dagger Y_\nu$ can be much smaller
than its diagonal elements. On the other hand, the dependence of
$b_H$ on $B_\nu$ is through $\sum_{ki} |(Y_\nu)_{ki}|^2$ which is
larger than the maximum $|(Y_\nu)_{ki}|^2$. Thus, bounds from the
LFV rare decay discussed in \cite{khodam} and from the condition
of electroweak symmetry breaking discussed in this paper are
complementary.

The phase of $B_\nu$ is a source of CP-violation and it can induce
electric dipole moment for charged leptons. At one-loop level, the effect
is suppressed by inverse powers of $M$ \cite{tamar}. However, at the
two-loop level
there is not such a suppression:
The neutrino $B$-term gives a correction to the $A$-term of the charged
leptons, $A_\ell$, independent of $M$ \cite{khodam}. Then, the imaginary
part of $A_\ell$ [which,
in the case
Im$(a_0)=0$ is given by Im$(B_\nu)$] gives a correction to the EDMs of
charged leptons \cite{khodam} while the real part gives a contribution to
the MDM of
charged leptons \cite{gilad}.
As in the case of LFV rare decays, we can argue that for $B_\nu
Y_\nu^2/(16
\pi)^2\gg m_{susy}$, these effects are suppressed because  the mass
of $\tilde{L}$ becomes too large. The possibility of such large
$|B_\nu|$ is ruled out
by the bound that we discussed in the previous section so the bounds on
the EDMs of charged leptons can be translated into bounds on
Im$(B_\nu)$ without this ambiguity. The present  bound on the electric
dipole moment of the electron
($d_e<1.4\times 10^{-27} $ e~cm \cite{pdg})
implies ${\rm Im}(B_\nu)\sum_i |(Y_\nu)_{ie}|^2/(16\pi^2)< m_{susy}$
which is again complementary to the bound we found in the previous
section.

All these bounds are summarized in Table 1. Notice that in the near future
the
bounds on $d_e$ and the branching ratios of the LFV rare decays
will be dramatically improved.
\section{Summary}
In this paper, we have first proposed a simple model in which
$B_\nu \gg m_{susy}$, introducing a new singlet chiral superfield, $X$.
The model is very similar to the singlet Majoron
model with the difference that the $F$-component of $X$ also develops a
VEV.

We have then shown that $B_\nu$ gives  corrections to the Higgs mass
parameters $b_H$ and $m_{H_u}^2$. We have discussed that, to satisfy the
condition for the electroweak symmetry breaking, $|b_H-B_\nu \mu {\rm
Tr}[Y_\nu Y_\nu^\dagger ]/(16 \pi^2)|$ has to be of order of
$m^2_{susy}/\tan \beta$ and, as a result, $B_\nu \sum_{ki}
|(Y_\nu)_{ki}|^2$
cannot be much larger than $16 \pi^2 m_{susy} /\tan \beta$ [see Eq.
(\ref{dis})].
We then have discussed how this bound complements our knowledge of this
sector, arguing that without this piece of information one could evade all
the bounds on $B_\nu$ that previously had been discussed in the literature
\cite{g-h,khodam}.
\section*{Acknowledgments}
I would like to thank B. Bajc, A. Brignole, A. Masiero, A. Rossi and G.
Senjanovic for useful discussion.
I am also very grateful to M. E. Peskin for his useful remarks.
The early stage of this work was supported by DoE under contract
DE-AC03-76SF00515.
 
\newpage

 \begin{table}
\begin{center}
\begin{tabular}{|c|c|} \hline
Neutrino mass  & $|B_\nu|<10^{3}m_{susy}$
\cite{g-h} \\
\hline
Br($\mu \to e \gamma$) \ \  & Re$(a_0^*
B_\nu)(Y_\nu^\dagger Y_\nu)_{\mu e}/(16 \pi^2)<
10^{-4}m_{susy}^2$  \cite{khodam} \\
\hline Br($\tau \to e \gamma$) \ \  & Re$(a_0^*
B_\nu)(Y_\nu^\dagger Y_\nu)_{\tau e}/(16 \pi^2)< 0.1 m_{susy}^2$
\cite{khodam}\\
\hline
Br($\tau \to \mu \gamma$)  & Re$(a_0^*
B_\nu)(Y_\nu^\dagger Y_\nu)_{\tau \mu}/(16 \pi^2)<
0.01 m_{susy}^2$  \cite{khodam}\\ \hline
$d_e <1.4 \times 10^{-27}$ e cm & Im($B_\nu)(Y_\nu^\dagger
Y_\nu)_{ee}/(16 \pi^2)<0.1
m_{susy}$ \cite{khodam}\\
\hline Electroweak symmetry breaking &
$|\mu B_\nu|{\rm Tr} [Y_\nu Y_\nu^\dagger ]/(16\pi^2)< m_{susy}^2/\tan \beta$ \\
\hline
\end{tabular}
\baselineskip 0.1 cm
\caption{A summary of different bounds on $B_\nu$. For simplicity, we have
assumed
that all the supersymmetry breaking parameters except $B_\nu$ are given
by $m_{susy}\sim 200$ GeV. The analytical formula in the text provide the
exact
dependence on the different susy breaking parameters.  }
\end{center}
\end{table}


\begin{thebibliography}{99}
\bibitem{gilad}
I would like to thank Gilad Perez for drawing my attention to this point.
\end{thebibliography}
\end{document}